\documentclass[aps,prl,reprint,superscriptaddress,amsmath,amsfonts,amssymb]
{revtex4-1}

\usepackage{graphicx}
\usepackage{hyperref}
\usepackage{amsthm}

\newcommand{\uba}{Departamento de F\'\i sica, FCEyN, UBA, Pabell\'on 1,
  Ciudad Universitaria, 1428 Buenos Aires, Argentina}
\newcommand{\ifiba}{Instituto de F\'\i sica de Buenos Aires, UBA CONICET,
  Pabell\'on 1, Ciudad Universitaria, 1428 Buenos Aires, Argentina}
  
\newcommand{\berk}{Department of Statistics, University of California, Berkeley, 367 Evans Hall, CA 94720, USA }

\newtheorem*{proposition}{Proposition}

\newcommand{\sys}{\mathcal S}
\newcommand{\reservoir}{\mathcal B}
\newcommand{\battery}{\mathcal W}
\newcommand{\clock}{\mathcal C}
\newcommand{\rev}{\sigma}

\newcommand{\Thot}{T_\text{hot}}
\newcommand{\Tcold}{T_\text{cold}}
\newcommand{\Bhot}{\beta_\text{hot}}
\newcommand{\Bcold}{\beta_\text{cold}}
\newcommand{\kB}{k_B}
\newcommand{\kBT}{k_BT}

\newcommand{\Wform}{W_{\text{form}}}
\newcommand{\Wextr}{W_{\text{ext}}}
\newcommand{\DF}{\Delta F}
\newcommand{\Wcycle}{W_{\text{cycle}}}
\newcommand{\Qhot}{Q_\text{hot}}
\newcommand{\Qcold}{Q_\text{cold}}
\newcommand{\etaCarnot}{\eta_\text{Carnot}}

\newcommand{\Hi}{H_1}
\newcommand{\Hf}{H_2}
\newcommand{\Gap}{\omega}
\newcommand{\Gapi}{\omega_1}
\newcommand{\Gapf}{\omega_2}
\newcommand{\fluctW}{\Delta W}

\newcommand{\Suppi}{\mathcal{U}}
\newcommand{\Suppf}{\mathcal{V}}
\newcommand{\restr}[2]{{#1}|_{#2}}

\newcommand{\rhoNcorr}{\rho^{(N)}}
\newcommand{\rhoNprod}{\rho^{\otimes N}}
\newcommand{\pbeta}{p_{\beta}}
\newcommand{\pbetahot}{p_{\Bhot}}
\newcommand{\pbetacold}{p_{\Bcold}}

\newcommand{\tr}{\text{tr}}
\newcommand{\ket}[1]{\left\vert{#1}\right\rangle}

\newcommand{\ketbra}[2]{|{#1}\rangle\langle{#2}|}
\newcommand{\dyad}[1]{\left\vert{#1}\middle\rangle\middle\langle{#1}\right\vert}
\newcommand{\comm}[2]{\left[{#1},{#2}\right]}
\newcommand{\Order}{\mathcal{O}}
\newcommand{\supp}{\text{supp}}
\newcommand{\comb}[2]{{{#1}\choose{#2}}}
\newcommand{\relent}[2]{D\left({#1}\middle\|{#2}\right)}
\newcommand{\Naturals}{\mathbb{N}}
\newcommand{\order}[1]{\mathcal{O}\left({#1}\right)}
\newcommand{\expval}[1]{\left\langle{#1}\right\rangle}
\newcommand{\beq}{\begin{equation}}
\newcommand{\eeq}{\end{equation}}
\newcommand{\bse}{\begin{subequations}}
\newcommand{\ese}{\end{subequations}}
\newcommand{\bea}{\begin{eqnarray}}
\newcommand{\eea}{\end{eqnarray}}

\usepackage{xcolor}

\begin{document}

\title{Heat engines with single-shot deterministic work extraction}

\author{Federico Cerisola}
  \email{cerisola@df.uba.ar}
  \affiliation{\uba} \affiliation{\ifiba}
\author{Facundo Sapienza}
  \email{fsapienza@berkeley.edu}
  \affiliation{\uba} \affiliation{\berk}
\author{Augusto J. Roncaglia}
  \email{augusto@df.uba.ar}
  \affiliation{\uba} \affiliation{\ifiba}


\begin{abstract}
  We introduce heat engines working in the nano-regime that 
  allow to extract a finite amount of deterministic work. 
  We show that the efficiency of these cycles is strictly smaller than Carnot's, and we associate 
  this difference with a fundamental irreversibility that is present in single-shot transformations.
  When fluctuations in the extracted work are allowed there is
  a trade-off between their size and the efficiency. As the size of
  fluctuations increases so does the efficiency, and optimal efficiency is recovered
  for unbounded fluctuations, while certain amount of deterministic work is drawn from the cycle.  
  Finally, we show that if the working medium is composed by many
  particles, by creating an amount of correlations between the subsystems that
  scales  logarithmically with their number, Carnot's efficiency can also be approached 
  in the asymptotic limit along with deterministic work extraction.
\end{abstract}


\maketitle


\noindent\emph{Introduction.--} Since its formulation, thermodynamics has become one
of the cornerstones of physics. Originally motivated by the study of
macroscopic thermal machines like steam engines, it  has now been pushed well
outside its original scope into the limit of small number of systems in the quantum realm \cite{goold2016,cinjanampathy2016,binder2018}. 
Pursuing the identification of the limitations and advantages of these devices that operates 
in the nano-regime,  
an extensive deal of work has been devoted  to the  study for instance of 
cycles analogous to Carnot's \cite{scovil1959, geusic1967, kosloff1984, geva1992, geva1996, 
feldmann2000, tonner2005,dann2020quantum} or Otto's \cite{feldmann1996, feldmann2003, quan2007,
rossnagel2016, camati2019coherence, tacchino2020non}, the performance of quantum refrigerators \cite{kosloff2000, palao2001, linden2010,
levy2012}, heat engines that exploit the quantumness 
nonclassical reservoirs \cite{rossnagel2014, brunner2014, scully2003} or
quantum measurements \cite{biele2017, elouard2017, elouard2018, ding2018,
buffoni2019}. Like in the standard scenario,
most of these analyses were focused on the study of average work extraction.
This assumption is well justified in the macroscopic limit due to the fact that the
amount of fluctuations decreases with the number of particles. 
However, in small systems work fluctuations dominate and may be even greater than the mean value of work.
Therefore, it becames relevant  to understand limitations of heat engines in single realisations
with controlled, or bounded, fluctuations of work in this regime.


Among the different approaches that have been developed to characterize
single-shot  thermodynamic transformations of nanoscale systems in contact with a thermal bath, 
a recent framework that gained a lot of interest is the so-called resource theory of
thermodynamics \cite{janzing2000, brandao2013, aaberg2013, horodecki2013, brandao2015,
lostaglio2015b, egloff2015, halpern2015, halpern2016, richens2016,
cwiklinski2015, gemmer2015, kwon2018, lostaglio2015c, lostaglio2017,
lostaglio2018, masanes2017, mueller2017, narasimhachar2015, ng2015,
sparaciari2017, vandermeer2017, sparaciari2017}. 
Within this framework,
a detailed account of every energy exchange between system and heat bath 
imposes severe restrictions to the allowed thermodynamic transformations
that go beyond the standard second law~\cite{aaberg2013, horodecki2013, brandao2013,brandao2015}.  
In fact, this set of restrictions determines that, in general, the minimum amount of deterministic 
work yielded in a given transformation is greater than the maximum work that can be 
drawn from the reverse process~\cite{horodecki2013, aaberg2013,brandao2013,brandao2015}.
Remarkably, the emergence of this fundamental notion of irreversibility 
is absent in the standard scenario where the free energy difference determines 
either the work that can be extracted from a given transformation, or the work that needs to be invested
to generate it. Thus, naturally, one expects a poorer performance for heat engines working in such 
regime. However, the existing results seems suggest that it is not even possible to 
design a cycle able extract a finite amount of deterministic work  in the single-shot regime \cite{brandao2013, richens2016, ng2017, woods2019}.

Here we show that in fact one can define such heat engines in the single-shot regime.
Specifically, we introduce thermodynamic cycles that allow to extract a deterministic amount of work from a nanoscale system (working medium) that  is in contact with two thermal baths.
These cycles can be described in two ways: either in terms of the collection of equilibrium states 
that the working medium reaches at the end of each stroke when
is subjected to a driving, or in terms of a set of non-equilibrium
states through which the working medium passes after the different strokes with fixed Hamiltonian. 
We show that the efficiency of these engines is strictly smaller than Carnot for determinist work extraction. 
These two types of engines also allow us to analyze the influence of fluctuations of work and 
the size of the working medium on the efficiency.
Indeed, we show that the efficiency of these engines can be 
enhanced either by allowing fluctuation in the extracted work or by increasing the size
of the working medium.\\


\begin{figure}[tb]
 \centering
  \includegraphics[width=0.45\textwidth]{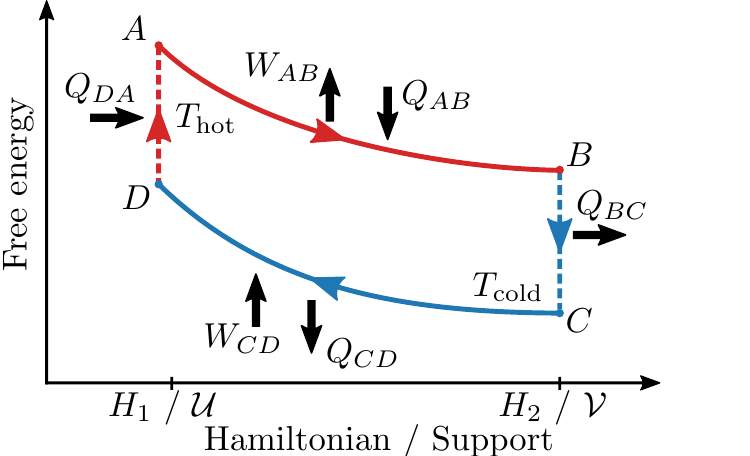}
  \caption{\label{fig:cycle}
    Pictorial representation of thermodynamic cycles for deterministic work extraction.
    For the cycles connecting equilibrium states, the transformation
     $A \to B$ or $C \to D$ are defined through the initial and final Hamiltonians.
     For the engines operating between non-equilibrium
    states, the states at $A$ and $B$ are identified by the support on energy eigenbasis, 
    which uniquely determines these states. 
    The strokes $A \to B$ and $C \to D$ are reversible processes even in
    the single-shot regime, while the two other strokes involve an irreversible
    thermalization. 
    }
\end{figure}

\noindent\emph{Single-shot scenario.--}
Let us start by setting up the scenario we will consider.
The fundamental components of an ordinary heat engine are two thermal baths at
different temperatures $\Thot$ and $\Tcold$ (with $\Tcold < \Thot$) and a
working medium $\sys$ that undergoes a cyclic transformation. 
We  assume that the working medium is an arbitrary finite dimensional quantum system,
and we have at our disposal two infinite heat baths in thermal states $\tau_\reservoir$. 
Work will be quantified by considering an additional degree of freedom, 
that acts as a battery \cite{aaberg2013, horodecki2013}.
The battery is modelled as quantum system $\battery$ with its own
Hamiltonian $H_\battery$ that, in a deterministic (fluctuation free) transformation, 
starts and ends up in a pure energy eigenstate of $H_\battery$. Thus, if the initial 
state of the battery is an eigenstate with energy $E_1$, and after a given transformation
it ends up in an eigenstate with energy $E_2$, we will then say that 
an amount of deterministic work $W = E_2 -E_1$ (if $E_2 > E_1$) has been drawn 
or it has been yielded (if $E_2 < E_1$) in the transformation. Thus, for deterministic
work extraction, a two-level system will be enough.

\begin{figure*}[tb]
  \includegraphics[width=0.9\textwidth]{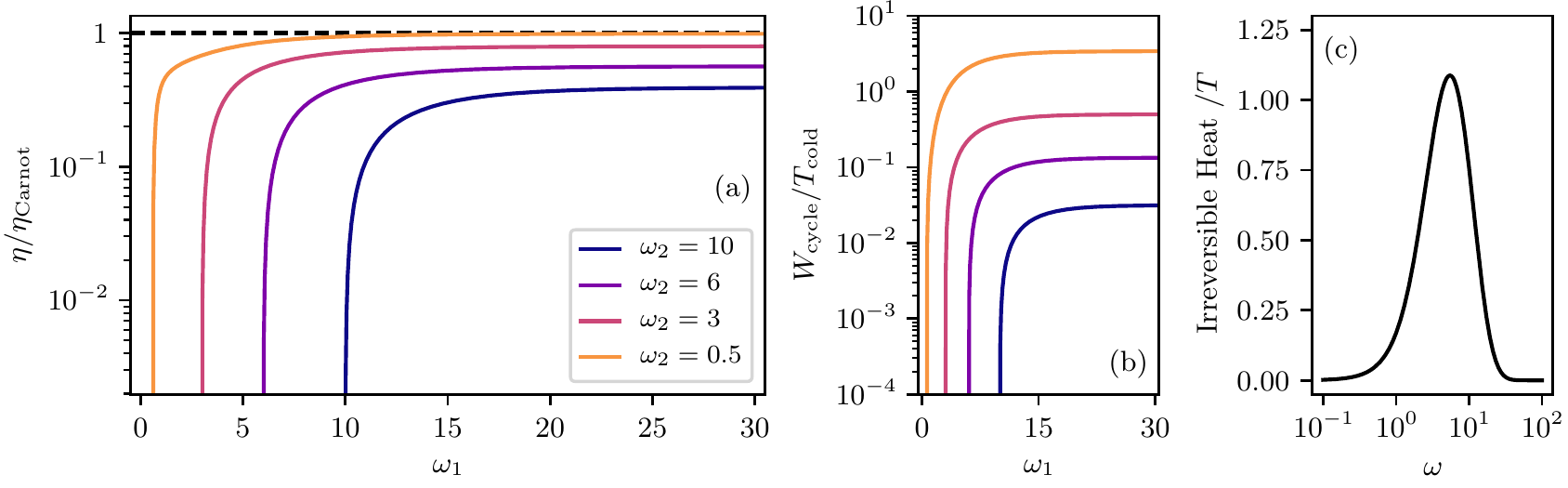}
  \caption{\label{fig:effeqrevlimq}
    Efficiency, work and irreversible heat for a single qubit heat engine of the first type. 
    $\Gapi$ and $\Gapf$ are the gaps of the initial and final Hamiltonians respectively, $\hbar=1$,
    $\kB=1$ and $\Tcold = 1$ in arbitrary units so that all
    energy quantities are expressed in units of $\Tcold$.
     (a) the efficiency of the cycle, 
    as $\hbar\Gapf \ll \kB\Tcold$ and $\hbar\Gapi \gg \kB\Thot$ Carnot's
    efficiency is approached.  (b) deterministic work drawn from the cycle. (c) irreversible heat 
    dissipated during the thermalizations steps (either $B \to C$ or $D \to A$) 
    as a function of the corresponding gap $\Gap$ {($\Gapi$ or $\Gapf$)}. 
  }
\end{figure*}

The cycles we will introduce can be generically described by a four stroke process
as it is depicted in Fig.~\ref{fig:cycle}.
At the beginning of each cycle the system, battery
and baths start uncorrelated in a product state,
and during each stroke the system interacts with only one of the baths.
As we are interested in work extraction in small systems we will take into account
all sources of energy transfer \cite{aaberg2013, horodecki2013,brandao2015}. 
Thus, we also assume that during each stroke the components interact 
through an energy-preserving unitary transformation $U$,
such that $\comm{U}{H_\sys + H_\battery + H_{\reservoir} } = 0$,
where $H_\sys$ is the Hamiltonian of the systems $\sys$ and $H_{\reservoir}$ is the Hamiltonian of the 
corresponding bath $\reservoir$. This is a strict energy conservation requirement, analogous
to the first law, and ensures that the unitary transformation is not injecting energy. Thus,
all energy exchanges with the battery come from the bath and/or the working medium.
In this way, an initial state  $\eta$ (system + battery) can be transformed
into a final one $\sigma$ after tracing over the degrees of freedom 
of the bath: $\sigma=\tr_{\reservoir}[U(\eta\otimes\tau_{\reservoir}) U^\dagger]$,
this is called a \emph{thermal transformation} and will be denoted by $\eta\rightarrow\sigma$~ \cite{horodecki2013,brandao2015}.

In this way,  each stroke can be generically described by a thermal transformation.
For a finite dimensional system with Hamiltonian $H_\sys=\sum_{E} E \,\Pi_E$, 
where $\Pi_E$ are projectors over the energy subspace $E$ in an initial block-diagonal state  
${\rho=\sum_{E,g} \lambda_{E,g}\dyad {E,g}}$ ($g$ accounts for the degeneracy of the energy levels) the maximum amount of deterministic work $\Wextr$ that can be
extracted in contact with a reservoir at temperature $T$ is given by
\begin{equation} \label{eq:wextr}
  \Wextr(\rho) =
   F_0(\rho) - F(\tau_\sys),
\end{equation}
where $F_0(\rho)=-\beta^{-1}\log\sum_{E \in \supp(\rho)} e^{-\beta E}$, 
$\beta = (\kBT)^{-1}$ with $\kB$ the Boltzmann constant,
 $\supp(\rho)$ is the support of the state $\rho$, and $F(\tau_\sys)$ is the standard
free energy of the thermal state $\tau_\sys=e^{-\beta H_\sys}/Z_\sys$ given by $F(\tau_\sys) = -\log Z_\sys$. This is called the \emph{extractable work}, and 
is obtained by maximizing $W$ over the thermal transformation 
$\rho \otimes \ketbra{0}{0}_\battery \rightarrow  \tau_\sys \otimes\ketbra{W}{W}_\battery$, with 
$H_\battery=W\ketbra{W}{W}_\battery$.
Notice that to be able to extract a nonzero deterministic amount  of work, the state cannot have full support
in the energy eigenbasis. 
The inverse transformation, where an 
non-equilibrium state $\rho$ is created out of an initial thermal state, 
requires a minimum amount of deterministic
\begin{equation} \label{eq:wform}
  \Wform(\rho) =   F_\infty(\rho) - F(\tau_\sys),
\end{equation}
where $F_\infty=\beta^{-1}\log\max_{E,g}\{{\lambda_{E,g} e^{\beta E}}\}$ which is the so-called \emph{work of formation}. Remarkably, ${\Wform(\rho) \geq \Wextr(\rho)}$ (the inequality is strict
except for very specific cases that we will discuss later) which means there is
a fundamental irreversibility in the single-shot regime \cite{horodecki2013,
brandao2015}.
Under all these assumptions, we will consider limits for microscopic heat engines. 



\emph{Cycles in terms of equilibrium states.-}
As it is usual in standard thermodynamics we will start by defining cycles in terms of thermal 
states of the working medium. In this way, one can notice from Eq.~\eqref{eq:wextr} that a stroke
with a fixed Hamiltonian starting from an initial thermal state is useless for deterministic work extraction, 
since no work can be extracted from full rank states. However, we can overcome this 
issue by introducing a driving, meaning that during the transformation
the Hamiltonian of the system changes from $\Hi$ to $\Hf$. 
This thermal transformation can be modelled
by introducing an auxiliary two-level  system $\clock$ with trivial Hamiltonian
that acts as a \emph{clock} \cite{horodecki2013, brandao2015}. Then, by defining the
Hamiltonian of the working medium as $H_{\sys\clock} = \Hi\otimes\dyad{0} +
\Hf\otimes\dyad{1}$, with $\{\ket{0}, \ket{1}\}$ an orthonormal basis of
$\clock$, the above work extraction process can be formally expressed as 
\begin{equation} \label{eq:clocktransform}
  \tau_{\sys,1}\otimes\dyad{0}_\clock\otimes\dyad{0}_\battery \to
  \tau_{\sys,2}\otimes\dyad{1}_\clock\otimes\dyad{W}_\battery,
\end{equation}
where $\tau_{\sys,i}$ is the thermal equilibrium state of $\sys$ with
Hamiltonian $H_i$. This type of transformation resembles the 
classical isothermal expansion of a gas.
It is easy to show now that the maximum 
deterministic work $W$ that can be extracted
 after this transformation is simply equal to the standard free energy
difference ${W = \DF = F(\tau_{\sys,2}) - F(\tau_{\sys,1})}$. Notably, this transformation 
holds an important property: it is \emph{reversible}, meaning that
the amount of work yielded in the inverse transformation is also equal to $W$.

\begin{figure*}[tb]
  \includegraphics[width=0.9\textwidth]{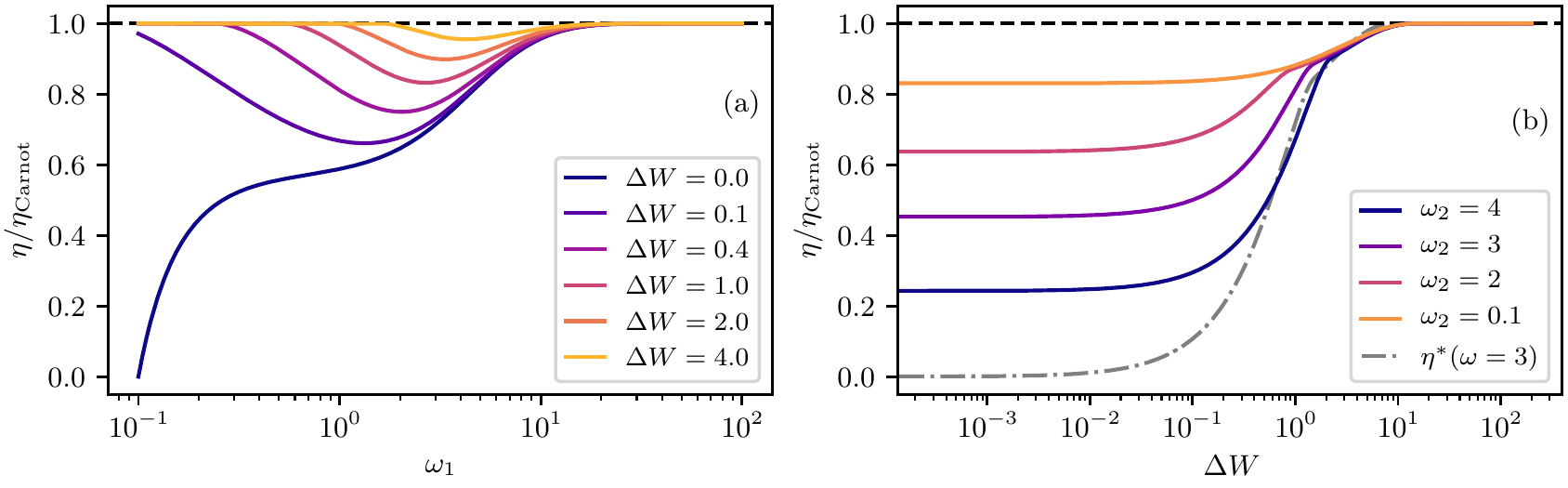}
  \caption{\label{fig:efffluctlim}
    Efficiency of a single qubit heat engine of the first type when fluctuations are allowed. 
    We use the same convention of Fig.~\ref{fig:effeqrevlimq} for numerics and  we fix $\Gapf = 5$.
     (a) shows the efficiency as a function of the initial gap $\Gapi$ for
    different fluctuation sizes $\fluctW$ (by
    this we mean that the extracted work must lay within $\expval{\Wcycle} \pm
    \fluctW$).
    Allowing fluctuations enhance the efficiency not only for
    large values of $\Gapi$ (as expected from Fig.~\ref{fig:effeqrevlimq}),
    but also for small ones, and as $\fluctW$ increases we
    eventually recover Carnot efficiency for all $\Gapi$. This can also be seen
    in panel (b) where it is shown the efficiency as a function of $\fluctW$ for
    different values of $\Gapi$. $\eta^*$ is the efficiency of the  single qubit heat engine proposed in        
    \cite{richens2016} working with a fixed Hamiltonian. For small values of $\fluctW$,
    the efficiency our cycle has a finite value while $\eta^*$ vanishes since it is not
     the cycle is unable to extract deterministic work.
  }
\end{figure*}

The four stroke cycles we define bellow ($A\rightarrow B \rightarrow C \rightarrow D \rightarrow A$) 
are illustrated in Fig.~\ref{fig:cycle}.
At the end of each stroke the system is in a thermal state, 
so we will label these states as $(H,T)$, indicating that the system is in equilibrium at temperature $T$ with Hamiltonian $H$. 
Let us analyse the cycle in detail, as we said before $\sys$ 
is a general finite dimensional system.
Initially, the system starts at $A$ in  equilibrium with $(\Hi,\Thot)$. 
The first stroke is such that  the system is driven from $\Hi\rightarrow \Hf$ in contact with the hot  bath  and ends up at $B$ in  equilibrium with $(\Hf,\Thot)$. 
The extracted work after this step is equal to $\DF_{AB} = F_A - F_B$  
(where  $F_{( \cdot )}$  is the standard free energy of the equilibrium state at each point).  
During the second stroke the system is brought in contact with the cold bath and 
thermalises, thus the state of the system at $C$  is $(\Hf,\Tcold)$. This transformation
is achieved at no work cost  and an amount of heat $Q_{BC}$ is dissipated in the cold bath 
(in terms of the resource theory this is consequence of the fact that the
thermal state is \emph{thermo-majorized} by all states \cite{horodecki2013}).
In the third stroke, at a work cost equal to $\DF_{DC} = F_D - F_C$, 
the system is driven from $\Hi\rightarrow \Hf$, still in contact with cold bath, and ends
up at $D$ in a thermal state $(\Hi,\Tcold)$.
Finally, the system is brought again in contact with the hot bath and thermalises after receiving
an amount of heat $Q_{DA}$, thus reaching the initial state.
In summary, after this cycle it is possible to extract a deterministic amount of work equal to:
\beq
  \Wcycle = F_A - F_B - F_D + F_C.
  \label{eq:wcycle}
\eeq
Notably, the derivation is general, as we did not impose any condition on the dimension of the system, 
Hamiltonians or temperatures. However, the work   $\Wcycle$ drawn in the  cycle of course depends on these details. 

The performance of any heat engine is evaluated by computing their efficiency.
In order to obtain the efficiency of these cycles, we have to compute 
the amount of heat exchanged  with the baths. 
If a thermal operation has an associated single-shot deterministic work cost $W$ and the
average internal energy change of the system is $\Delta E$, then  the
heat $Q$ exchanged with the reservoir during the transformation is 
\beq
Q= \Delta E - W.
\eeq 
Now, the efficiency is defined as the ratio between the extracted work and the heat exchange with the hot bath:
\begin{align}
  \eta
  &= \frac{\Wcycle}{\Qhot} = 1 - \frac{\Qcold}{\Qhot} \nonumber \\
  &= 1 - \frac{\Tcold\left[S_D - S_C\right] + E_C - E_B}
           {\Thot\left[S_A - S_B\right] + E_D - E_A},
  \label{eq:eff}
\end{align}
where $S_{(\cdot)}$ and $E_{(\cdot)}$ are the entropy and average energy
of the system in each state, respectively. One can easily check that this value is indeed 
strictly smaller than Carnot's efficiency,
$\eta < \etaCarnot = 1 - {\Tcold}/{\Thot}$.
As we will see, this is related with the
heat exchange during the thermalization processes $B \to C$
and $D \to A$, which are irreversible in the single-shot regime. 
In fact, while the transformation $B \to C$ 
can be done without investing work, the inverse transformation  
$C \to B$ requires a finite amount of work. This is due to the fact the state at $B$, $(\Hf, \Thot)$
is a non-equilibrium state for the cold bath. Hence, one can show that the efficiency can be 
improved as the heats $Q_{DA}$ and $Q_{BC}$ are reduced. In fact, as
$Q_{BC} \to 0$ and $Q_{DA} \to 0$,   $\eta \to \etaCarnot$. 
This behavior is illustrated in Fig.~\ref{fig:effeqrevlimq} for a single-qubit heat engine. 
There, the Hamiltonians at $A$ and $B$ are
 $\Hi = \hbar\Gapi\dyad{1}$ and $\Hf = \hbar\Gapf\dyad{1}$ respectively, where
$\{\ket{0}, \ket{1}\}$ is an orthonormal basis of $\sys$. 
In Fig.~\ref{fig:effeqrevlimq} we can see the efficiency, work and irreversible
heat exchange for this engine, thus for $\hbar \Gapf \ll k_{\rm{B}}\Tcold$ and $\hbar \Gapi \gg k_{\rm{B}}\Thot$ 
 the irreversible heat is drastically reduced and 
Carnot efficiency is approached. 
It is also worth noting that this cycle is very reminiscent of the
Stirling engine \cite{organ1997}. 
This classical cycle also contains irreversible thermalization
steps that reduce the efficiency. However, by adding an auxiliary system, typically called regenerator,
reversibility and Carnot's efficiency could in principle be recovered \cite{organ1997}. 

Interestingly, there is another way to approach Carnot's efficiency  in this cycle.
This is at the expense of allowing some fluctuations in the work extraction. 
Extensions of  single-shot thermodynamics with bounded fluctuations in work has been thoroughly
studied in \cite{richens2016}, and it was shown that if arbitrarily large
fluctuations are allowed it is possible to extract an \emph{average work} equal to the free energy difference.
In particular, this means that we can also extract some fluctuating work during 
the thermalization steps, 
 and if we allow arbitrary large 
fluctuations the mean value of this work equals the free energy difference: $\expval{W_{BC}} = F_B - F_C$
and $\expval{W_{DA}} = F_D - F_A$. It is then straightforward to see that in this limit the
efficiency is precisely Carnot, $\eta = \etaCarnot$. Notably, by allowing fluctuations we do not
chagne the deterministic work that is being drawn from
$A \to B$, since over this stroke work is already equal to the free energy
difference. Thus, if the size of fluctuations is bounded,
the average work that can be extracted during the thermalization step is $\expval{W}<\DF$, and
there is an improvement in the efficiency. 
 For a detailed analysis, a closed form of the average work
with bounded fluctuations can be obtained for two-level 
 systems \cite{richens2016}. In Fig.~\ref{fig:efffluctlim} we show the efficiency 
 as a function of the size of the fluctuations, $\fluctW$, for the single qubit heat engine. There we can see 
the efficiency increases as we allow fluctuations, and for large fluctuations
Carnot efficiency is approached. Furthermore, even for a small amount of fluctuations
the efficiency is drastically improved. \\

\noindent\textit{Cycles in terms of non-equilibrium states.--}
We will now show a cycle that generalizes the previous one, and can be defined in terms
non-equilibrium states with a fixed Hamiltonian. 
Deterministic work extraction with a fixed Hamiltonian requieres an initial 
non-full rank state for  $\sys$, and at the end of the cycle it is required the creation of this 
non-equilibrium state. However, since in general the work of formation is larger
than the extractable work, this poses an obstacle. To see how we can circumvent 
this issue we will introduce the family of reversible states~\cite{sapienza2019}.
We will say that a state $\rev$ is \emph{reversible} if its work of formation equals its extractable work: 
\begin{equation}
{\Wform(\rev) = \Wextr(\rev)}.
\end{equation}
The following results provides a complete characterization of this set of states. 

\begin{proposition}
Consider a block-diagonal state $\rev$ with support 
\begin{equation}
    \supp (\rev) = \bigoplus_{E} \mathcal U_E,
\end{equation}
where $\mathcal U_E$ is a subspace of the energy shell $E$. Then, $\rev$ is a reversible at a background inverse temperature $\beta$ if and only if has a thermal-like distribution over $\supp (\rev )$, i.e. 
\begin{equation}
    \rev = \frac{1}{Z} \sum_E e^{-\beta E} \, \Pi_{\mathcal U_E},
  \label{eq:revdefi}
\end{equation}
where $\Pi_{\mathcal U_E}$ is the projector over $\mathcal U_E$, and $Z$ is a normalization constant given by $Z = \sum_{E} \rm{dim}(\mathcal U_E) e^{-\beta E}$.
\label{prop}
\end{proposition}

Notice that these states have a uniform distribution over each $\mathcal U_E$.
Some key properties of reversible states
(which we prove in the Supplementary Material) are listed bellow:
\begin{enumerate}
  \item The work of formation and the extractable work of the reversible states are
    equal to the standard free energy difference 
    $$\Wform(\rev) =
    \Wextr(\rev) = F(\rev) - F(\tau(\beta)), $$
    where $\tau(\beta)$
     is the thermal state at inverse temperature of the bath $\beta$.
  \item Any state with the same support of $\rev$ can be 
    transformed into $\rev$ via a single-shot thermal operations at no work
    cost. \label{prop:revisthermomaj}
  \item Any state $\rho$ with the same support of $\rev$ and $\rho \neq \rev$, has
    $\Wextr(\rho) = \Wextr(\rev)$ and $\Wform(\rho) > \Wform(\rev)$.
\end{enumerate}

A certain family of the reversible states that has an interesting physical interpretation~\cite{sapienza2019},
and we will use bellow,  corresponds to the case where  energy level is uniformly populated or not populated at all, that is, $\text{dim} (\mathcal U_E) = 0$ or $\text{dim} (\mathcal U_E) = g(E)$, with $g(\cdot)$ the degeneracy. Each state of this set is fully characterized by the set of energies $\mathcal U$ that define their support, and can be written as:
\begin{equation}
  \restr{\tau}{\mathcal U}(\beta) =
  \frac{1}{Z_{\mathcal U}} \sum_{E \in \mathcal U} e^{-\beta E} \Pi_E.
  \label{eq:thermalsupp}
\end{equation}

Now we are ready to introduce the second set of heat engines, 
illustrated in Fig.~\ref{fig:cycle}. We will consider the system $\sys$ as an 
arbitrary finite dimensional system (as it will become clear later we cannot define a non-trivial single qubit heat engine for this cycle) with a given Hamiltonian. 
Without loss of generality (see the SM) we will assume that initially is in an non-equilibrium 
reversible state (i.e. not a thermal state) at inverse temperature $\Bhot$. 
During the first stroke, $\sys$ goes from 
$\restr{\tau}{\Suppi}(\Bhot)$ to $\restr{\tau}{\Suppf}(\Bhot)$  in contact with the hot bath.
As we mentioned, these two states are completely determined 
by their respective supports ($\Suppi$ and $\Suppf$) and
temperature. Since the initial and final states are reversible, 
the total amount of deterministic work that is drawn in this step equals the standard free energy difference: 
$W_{AB} = F_A - F_B =
F(\restr{\tau}{\Suppi}(\Bhot)) - F(\restr{\tau}{\Suppf}(\Bhot))$.
The second stroke, $B \to C$, is such that the system goes from
$\restr{\tau}{\Suppf}(\Bhot) \to \restr{\tau}{\Suppf}(\Bcold)$ in contact with the cold bath. 
This step generalizes the thermalization stroke of the previous cycle and, like in that case,
it can be achieved at no work cost  (see Property 2).  The remaining
strokes are defined in a similar way. 
During $C \to D$, the system is in contact with the cold bath
and  the transformation $\restr{\tau}{\Suppf}(\Bcold) \to \restr{\tau}{\Suppi}(\Bcold)$ is done at a deterministic
work cost equal to the free energy difference $W_{DC} = F_D- F_C$. Finally, the transformation $D \to A$, 
where the system returns to its initial state $\restr{\tau}{\Suppi}(\Bcold) \to \restr{\tau}{\Suppi}(\Bhot)$,
is done in contact with the hot bath at no work cost.
Therefore, the expressions for the net extracted work and the efficiency have the
same form as before (Eqs. \eqref{eq:wcycle} and \eqref{eq:eff}) except that in this case the
labels $A$, $B$, $C$, $D$ refers to the non-equilibrium states we define above. 
Notice that the previous cycle can also be considered a 
particular realization of this more general cycle. 
Indeed, when one adds the clock degree of freedom,
the complete state of system plus clock can be considered as
particular instances of the reversible states we defined before.

This more general cycle is particularly useful to analyse the behaviour of these heat
engines when the working medium $\sys$ is composed by $N$ identical subsystems.
In \cite{sapienza2019}, the presence of correlations in single-shot transformations was studied. 
In particular, it was shown that 
for every single particle state $\rho$ there exists a correlated $N$-partite state $\rhoNcorr$, such that 
the reduced state of each subsystem is $\rho$ and $\Wform(\rhoNcorr)\leq N\Wform(\rho)$. 
Notably, the set of reversible states of Eq.~\eqref{eq:thermalsupp}
appears naturally in this context, when one considers the set of states 
that minimizes the corresponding work of formation. Interestingly, if
we consider heat engines between these reversible states, 
 it is possible to demonstrate that for large $N$ the efficiency converges to Carnot.
As it was shown, the amount of correlations present in these
states scales as $\Order(\log N)$. Therefore, if we consider a working medium
composed of $N$ particles, we recover Carnot efficiency allowing an amount correlations
per particle that it is vanishing small in the large $N$ limit (see Supplementary Material).
Furthermore, one can show that in the limit $N\to\infty$,
$W(\rho^{(N)})/N \to \DF = F(\rho) - F(\tau)$, so that the extracted work 
per particle in this cycle is  simply given by the standard free energy of
the reduced state.



\textit{Discussion.--}
It is worth comparing our results with previous single-shot proposals that were unable to extract
deterministic work. In \cite{horodecki2013} a single-shot engine that mimics the Carnot cycle was introduced,
it consisted of two strokes in contact with heat baths plus two adiabatic transformations.
Our results indicate that if one replaces the adiabatic strokes with 
thermalizations (at no work cost), single-shot deterministic work extraction can be achieved.
A qubit heat engine with two strokes was introduced in \cite{richens2016}. There  
the transformations in contact with the heat baths were done at fixed Hamiltonian, and
it was shown that no deterministic work can be extracted. However,  when
fluctuations in work are allowed a non-zero average work can be extracted at finite efficiency.
Here we showed that deterministic work extraction with fixed Hamiltonian requieres strokes with
non-equilibrium states. In Fig.~\ref{fig:efffluctlim}-(b) we plot the efficiency $\eta^*$ of 
the heat engine introduced in \cite{richens2016} along with 
the efficiency of our first cycle for a two-level system.
Besides the difference in efficiency, we should also stress that in our cycle 
the amount of deterministic work does not change
when fluctuations are allowed, in fact the efficiency is improved because 
an additional fluctuating work that is extracted during the thermalization stroke.
Finally, in \cite{woods2019, ng2017} it was shown that no deterministic heat engine exists if the cold
bath has finite size. This is an interesting approach while is different from the one considered 
here and in the other proposals. Our scheme requires infinite hot
and cold baths, which is very much in line with traditional formulations of
heat engines.


We have introduced thermodynamic
cycles that allow deterministic work extraction in the single-shot regime. 
While previous work seem to suggest that it is not possible to define such
an engine, here we show some general deterministic cycles working with equilibrium and non-equilibrium states.
It is worth noting that while we have focused on an engine that extracts work,
the same idea can be used to design a single-shot refrigerator that has a
deterministic work cost of operation (although the heat removed from the cold
bath will still have fluctuations). Indeed both strokes, $A \to B$ and $C\to D$, 
are reversible and therefore can be inverted, while the
thermalization steps $B \to C$ and $D \to A$ are irreversible. Therefore to
operate as a refrigerator these transformations have to be changed.
However, it is easy to check that swapping the baths at those
steps (so that $C \to B$ is done at $\Thot$ and $A \to D$ at
$\Tcold$) is enough. 
We have also show that optimal efficiency can be approached by allowing fluctuations 
in the extracted work, or in the limit $N\to\infty$ when the working medium is composed of many particles.
In this last example, the cycle is such that the work extracted per particle depends only on the
standard non-equilibrium free energy of the reduced
system (which can be chosen arbitrarily). \\


\begin{acknowledgments}
  FC and AJR acknowledge support from CONICET, UBACyT and ANPCyT.
  We thank David Jennings for useful discussions about the connection between
  reversible states and Hamiltonian switching.
\end{acknowledgments}


\bibliography{biblio}

\newpage
\appendix


\section{Properties of reversible states}

\begin{proof}
(Proposition in the main text).
We will start by showing
that any state has has a thermal-like distribution over a reduced support is
reversible. Notice that all the non-zero eigenvalues $\lambda_{E, g}$ of the states $\sigma$ of the form \eqref{eq:revdefi} satisfy $\lambda_{E,g} e^{\beta E} = 1 / Z$. Then, 
\begin{align}
    \Wform(\sigma) 
    &= 
    \beta^{-1} \log \max_{E,g} \lambda_{E,g} e^{\beta E} - F(\tau(\beta)) \nonumber \\
    &=
    - \beta^{-1} \log Z - F(\tau(\beta)) \nonumber \\
    &= 
    - \beta^{-1} \log \sum_E \dim (\mathcal U_E) e^{-\beta E} - F(\tau(\beta)) \nonumber \\
    &= 
    \Wextr (\sigma),
\end{align}
which shows that $\sigma$ is a reversible state. 

Let us prove now that \textit{all} reversible states have the form $\eqref{eq:revdefi}$. Consider $\rho$ a reversible state and $\supp (\rho)$, with 
\begin{equation}
     \supp (\rho) = \bigoplus_{E} \mathcal U_E^\rho,
\end{equation}
where $\mathcal U_E^\rho = \supp (\rho) \cap \{ \ket{\psi} : H \ket{\psi} = E \ket{\psi} \}$. 
Let's define the thermal-like state $\sigma$ on the support of $\rho$, that is,
\begin{equation}
    \sigma = \frac{1}{Z_\sigma} \sum_E e^{-\beta E} \Pi_{\mathcal U_E^\rho}. 
\end{equation}
Since $\rho$ and $\sigma$ have the same support, they have the same extractable work, $\Wextr (\sigma) = \Wextr (\rho)$. 
Now both states
are reversible ($\rho$ by hypothesis and $\sigma$ we have
proven); therefore
\begin{equation}
  \Wform(\rho) = \Wextr(\rho) =
  \Wextr(\sigma) = \Wform(\sigma).
\end{equation}
This proves that both states also have the same work of formation. Based on the definition of work of formation, this implies 
\begin{equation}
    \max_{E,g} \lambda_{E, g}^\rho e^{\beta E} = \max_{E,g} \lambda_{E, g}^\sigma e^{\beta E} = \frac {1}{Z_\sigma},
\end{equation}
where $\lambda_{E,g}^\rho$, $\lambda_{E, g}^\sigma$ are the eigenvalues of $\rho$ and $\sigma$ with associated energy $E$, respectively. This last expression implies that $\lambda_{E, g}^\rho \leq e^{-\beta E} / Z_\sigma = \lambda_{E_g}^\sigma$. If the last inequality is strict for at least one value of $(E, g)$, then we will have 
\begin{equation}
    1 = \tr (\rho) = \sum_{E,g} \lambda_{E,g}^\rho < \sum_{E, g} \lambda_{E,g}^\sigma = 1,
\end{equation}
which is a contradiction. We conclude $\lambda_{E,g}^\rho = \lambda_{E, g }^\sigma$ and therefore $\rho = \sigma$. 
\end{proof}

Now we prove properties 1--3 of reversible states. Notice that since
$\sigma$ has a Gibbs distribution over the subspace
$\mathcal{U}$, then clearly the free energy is
\begin{equation}
  F(\sigma) =
    -\kBT \log \sum_E \dim (\mathcal U_E) e^{-\beta E_n}.
\end{equation}
Putting all this together means that $\Wform(\sigma) =
\Wextr(\sigma) = F(\sigma) - F(\tau)$.
This proves property 1 for this class of reversible states.

For property 2, given any state $\rho$ with support $\supp (\sigma)$ and such that
$\comm{\rho}{H} = 0$, we want to show that the transformation $\rho \to \sigma$
is allowed by thermal operations. Showing this is equivalent to showing that
$\rho$ \emph{thermo-majorizes} $\sigma$,
\cite{horodecki2013}. Now, given that both states have the same support, the
thermo-majorization condition is the same as that for two full-rank probability
vectors with the same distribution $\rho$ and $\sigma$ have
over $\supp (\sigma)$. Within this subspace $\sigma$ has a
thermal distribution and therefore it is majorized by \emph{any} vector with
the same support, in particular $\rho$.

For property 3, since $\rho$ and
$\sigma$ have the same support, they have the same
extractable work $\Wextr(\rho) = \Wextr(\sigma)$.
Furthermore, we have shown in property 2 that any state $\rho$ with the same
support as $\sigma$ can be converted into the latter at no
cost. Therefore the cost of formation of any state $\rho$ must be at least
that of $\sigma$, $\Wform(\rho) \geq \Wform(\sigma)$. Since the only reversible state with the same support than $\sigma$ is $\sigma$, we have that the last inequality is strict.

In light of these properties, if we have the system initially in an
arbitrary non-equilibrium state $\rho$ with non full support, we can always transform it to the
reversible state, $\restr{\tau}{\Suppi_\rho}$, with same support $\Suppi_\rho$.
In this way, one can extract the same amount of work from both states, although the reversible one
has a smaller work of formation. 
Analogously, we can easily see that if the system is initially in a thermal state, we can transform it to
an arbitrary reversible state using an amount of energy equal to its work of formation (this energy can
be recovered after some finite number of cycles).
Thus, without loss of generality we will consider that the system is initially in an non-equilibrium 
reversible state (i.e. not a thermal state). 


\section{Correlated subsystems}

Consider a working medium of $N$ non-interacting identical qubits. If the gap
of the qubits is $\hbar\Gap$, then the $N$ qubit system will have an energy
spectrum $\{E_m = m\hbar\omega, m = 0, \dots, N\}$, each with degeneracy $g(m)
= \comb{N}{m}$.
We will focus on states $\rhoNcorr$ of the $N$ qubits such that the local
density matrix of all qubits is the same. Given that we will restrict ourselves
to states diagonal in the energy eigenbasis, these local states can be
parametrized by the excited state probability $p$ as $\rho = p\dyad{1} +
(1-p)\dyad{0}$, where $\{\ket{0}, \ket{1}\}$ are an orthonormal basis such that
$H = \hbar\omega\dyad{1}$.

In \cite{sapienza2019} it was shown when such global states $\rhoNcorr$ of the
$N$ qubits with correlations are allowed, then one can have a global work cost
of formation lower than that of the uncorrelated subsystems, that is
\begin{equation}
  \Wform(\rhoNcorr) < \Wform(\rhoNprod) = N\Wform(\rho).
\end{equation}
In particular, for certain single qubit states $\rho$, the correlated state
$\rhoNcorr$ that minimizes $\Wform(\rhoNcorr)$ is a reversible state. This happens for example for local single-qubit states
$\rho_k = p_k\dyad{1} + (1-p_k)\dyad{0}$, $k = 1, \dots, N-1$,  such that
\begin{equation} \label{eq:pkrevdef:app}
  p_k = \frac
    {\sum_{m=0}^{k} m \comb{N}{m} e^{-m\beta\hbar\omega}}
    {N\sum_{m=0}^{k} \comb{N}{m} e^{-m\beta\hbar\omega}}.
\end{equation}
The respective reversible correlated $N$ qubit states have a Gibbs-like thermal
distribution over the support $\mathcal{U}_k = \{\ket{E = m\hbar\omega, g}, g =
1, \dots, g(m); m = 0, \dots, k\}$. All relevant
thermodynamic quantities of these states are determined by the effective
partition function $Z_k(\beta)$ given by
\begin{equation}
  Z_k(\beta) = \sum_{m = 0}^{k} \comb{N}{m}\,e^{-m\beta\hbar\omega}.
\end{equation}
For the large $N$ limit that we will be taking later on, it is useful to
rewrite this partition function as
\begin{equation} \label{eq:partfuncrev:app}
  Z_k(\beta) = \left[Z(\beta)\right]^N
    \sum_{m = 0}^{k} \comb{N}{m}\,\pbeta^{m}(1-\pbeta)^{N-m},
\end{equation}
where $\pbeta = e^{-\beta\hbar\omega}/(1 + e^{-\beta\hbar\omega})$ is the
excited state probability of a single qubit in thermal equilibrium and
$Z(\beta) = 1 + e^{-\beta\hbar\omega}$ its respective partition function.
Notice that the second term in \eqref{eq:partfuncrev:app} is a tail sum of a
binomial distribution characterized by $N$ trials with success probability
$\pbeta$. We will base our large $N$ approximation on well known approximations
for binomial tails.

Let $\{k_N \in \Naturals, N \in \Naturals_0\}$ be a sequence such that
$k_N/N \to q$, when $N \to \infty$, for some $0 < q < \pbeta$. We then have the
following bounds on the binomial tail $B(\pbeta, k_N, N) = \sum_{m = 0}^{k_N}
\comb{N}{m}\,\pbeta^{m}(1-\pbeta)^{N-m}$ (\cite{ash1990}, Lemma 4.7.2)
\begin{equation}
  \frac{1}{\sqrt{8Nq(1-q)}}
  e^{-N\relent{q}{\pbeta}}
  \leq
  B(\pbeta, k_N, N)
  \leq
  e^{-N\relent{q}{\pbeta}},
\end{equation}
where $\relent{q}{p} = q\log(q/p) + (1-q)\log((1-q)/(1-p))$ is the binary
relative entropy. Therefore asymptotically we have that \cite{ash1990}
\begin{equation}
  \lim_{N\to\infty} -\frac{1}{N}\log B(\pbeta, k_N, N) = \relent{q}{\pbeta},
\end{equation}
and the convergence rate is $\order{\log N / N}$. Applying this results to the
logarithm of the partition function \eqref{eq:partfuncrev:app} of the
reversible state we have the asymptotic behaviour
\begin{equation}
  -\frac{1}{N}\log Z_k(\beta) \approx
  -\log Z(\beta) + \relent{q}{\pbeta}.
\end{equation}
Therefore, the average energy per qubit is
\begin{align} \label{eq:avgenperp:app}
  \frac{1}{N}\expval{E}_k
  &= -\frac{1}{N}\frac{\partial}{\partial\beta} \log Z_k(\beta) \nonumber \\
  &\approx \pbeta\hbar\omega + \frac{\partial}{\partial\beta}
    \relent{q}{\pbeta} \\
  &= \pbeta\hbar\omega + q\hbar\omega - \pbeta\hbar\omega = q\hbar\omega.
\end{align}
Notice that the average energy of the system can also be written in terms of
the local state probability $p_k$ as $\expval{E}_k = Np_k\hbar\omega$.
Therefore, this implies that in the large $N$ limit $p_k \to q$, so that the
parameter $q$ determines the asymptotic local stat of the qubits. Here all
convergence rates are of order $\order{\log N / N}$. Similarly, for the free
energy of the reversible state we have that \cite{sapienza2019}
\begin{equation}
  F_k(\beta) = NF(q,\beta) + \order{\log N},
\end{equation}
where $F(q,\beta)$ is the standard free energy of the asymptotic single qubit
state $\rho(q) = q\dyad{1} + (1-q)\dyad{0}$. In \cite{sapienza2019} it is
further show that the total correlations per particle in these reversible
states vanish in the large $N$ limit as $\order{\log N / N}$.

In our heat engine cycle we need to have reversible states at two different
temperatures, $\Bhot$ and $\Bcold$, and different supports $\mathcal{U}$ and
$\mathcal{V}$ for each number of subsystems $N$. If we choose our sequence of
supports $\mathcal{U}_{k_N}$ and $\mathcal{V}_{l_N}$ such that $k_N / N \to q$
and $l_N / N \to r$ with $0 < q < r < \min(\pbetahot, \pbetacold)$, we can then
apply the above asymptotic expressions for all four reversible states. In
particular, this means that the work extracted in each cycle per particle is
\begin{equation}
  \frac{\Wcycle}{N} \approx
  F(q,\Bhot) - F(r,\Bhot) - F(q,\Bcold) + F(r,\Bcold),
\end{equation}
so that we simply extract the free energy difference of the local states.
Notice that this local states have full support and therefore it would be
impossible to deterministically extract any energy from them without the
correlations. Furthermore, notice that in the large $N$ limit we have that the
average energy per particle \eqref{eq:avgenperp:app} does not depend on
temperature. This means that the reversible states at points $B$ and $C$ of the
cycle (or $A$ and $D$) have, up to $\order{\log N / N}$ corrections, the same
energy. This implies that the irreversible heat per particle vanishes in the
large $N$ limit, $Q_{BC} \approx 0$, $Q_{DA} \approx 0$. As discussed in the
main text, this would imply that we have Carnot efficiency. Indeed it is simple
to check via direct subsitution of the above asymptotic expressions that
\begin{equation}
  \lim_{N\to\infty} \eta = \etaCarnot.
\end{equation}


\end{document}

